\documentclass[12pt]{article}
\usepackage{amssymb,amsmath}
\usepackage{graphicx}
\usepackage{epsfig}
\usepackage{epstopdf}
\usepackage{latexsym}
\usepackage{psfrag}
\usepackage{booktabs}
\usepackage{bbm}

\usepackage[toc]{appendix}

\usepackage{color}
\usepackage{datetime}
\usepackage[
      colorlinks=false,
      linkcolor=darkblue,  
      urlcolor=blue,    
      filecolor=blue,     
      citecolor=red,
linktocpage=true,
      pdfstartview=FitV,
      bookmarksopen=true    
      ]{hyperref}

\ifpdf
\fi

\renewcommand{\theequation}{\arabic{section}.\arabic{equation}}

\def\ds{\displaystyle}

\definecolor{cardinal}{rgb}{0.6,0,0}
\definecolor{darkgreen}{rgb}{0,0.5,0}
\definecolor{golden}{rgb}{0.92, 0.7, 0}
\definecolor{midnight}{rgb}{0, 0, 0.5}
\definecolor{darkblue}{rgb}{0.2, 0, 0.8}

\begin{document}  

\begin{titlepage}
 
\bigskip
\bigskip
\bigskip
\bigskip
\begin{center} 
{\Large \bf   Schr\"odinger Deformations of $AdS_3\times S^3$}

\bigskip
\bigskip 

{\bf Nikolay Bobev${}^{(1)}$  and Balt C. van Rees${}^{(2)}$ \\ }
\bigskip
${}^{(1)}$
Simons Center for Geometry and Physics\\
${}^{(2)}$
C.N. Yang Institute for Theoretical Physics\\
Stony Brook University\\
Stony Brook NY 11794-3840, USA\\
\bigskip
nbobev@scgp.stonybrook.edu,~~vanrees@insti.physics.sunysb.edu  \\
\end{center}

\vskip 20mm

\begin{abstract}

\noindent We study Schr\"odinger invariant deformations of the $AdS_3\times S^3\times T^4$ (or $K3$) solution of IIB supergravity and find a large class of solutions with integer and half-integer dynamical exponents. We analyze the supersymmetries preserved by our solutions and find an infinite number of solutions with four supersymmetries. We study the solutions holographically and find that the dual D1-D5 (or F1-NS5) CFT is deformed by irrelevant operators of spin one and two.

\end{abstract}

\end{titlepage}


\tableofcontents

\section{Introduction}

The various gauge/gravity dualities have proven to be a powerful tool in the investigation of strongly coupled quantum field theories. The theories which can be studied with the standard dualities are always relativistic. However, in condensed matter and atomic physics one also encounters quantum critical points described by \emph{non-relativistic} strongly coupled quantum field theories. A particular example of such critical points are quantum field theories with the so-called Schr\"odinger invariance \cite{Nishida:2007pj}. The Schr\"odinger algebra is the Galilean algebra amended with scale invariance and a non-relativistic special conformal transformation. The scale transformations, $t\to \lambda^{n} t$ and $x\to \lambda x$, act differently on space and time and therefore break relativistic invariance and the full conformal group. The real number $n$ is called the dynamical exponent. Strictly speaking the Schr\"odinger algebra is defined only for $n=2$ since for any other value of $n$ there are no special conformal transformations, see for example \cite{Adams:2008wt} for details\footnote{ In this paper, with slight abuse of terminology, we call the non-relativistic scale invariant algebra for any $n$ the Schr\"odinger algebra and hope that this will not cause confusion.}.

A first attempt to extend the realm of the gauge/gravity dualities to include Schr\"odinger invariant quantum field theories was made in \cite{Son:2008ye,Balasubramanian:2008dm}. In these papers Schr\"odinger invariant solutions with $n=2$ were found by studying an effective model of Einstein gravity with a negative cosmological constant coupled to a massive vector field. In addition the authors of \cite{Balasubramanian:2008dm} used holographic techniques to compute correlation functions in the dual field theory and justify that the gravity backgrounds are indeed dual to non-relativistic scale invariant quantum field theories. 

In \cite{Adams:2008wt,Herzog:2008wg,Maldacena:2008wh} the Schr\"odinger invariant solutions of \cite{Son:2008ye,Balasubramanian:2008dm} were embedded in type IIB supergravity by performing a particular deformation of the Freund-Rubin $AdS_5\times SE_5$ compactifications, where $SE_5$ is a Sasaki-Einstein manifold. Later it was realized that $AdS_5$ compactifications of type IIB supergravity admit a much larger class of Schr\"odinger invariant deformations, for integer and half-integer dynamical exponents, and the dynamical exponent is determined by an eigenvalue of the Laplacian on the compact Einstein manifold \cite{Hartnoll:2008rs,Donos:2009en,Bobev:2009mw,Donos:2009xc,Donos:2009zf}. Some of the solutions studied in these papers are supersymmetric.

Apart from possible applications as holographic duals of quantum critical points with non-relativistic scale invariance, Schr\"odinger invariant backgrounds are also interesting from a different perspective. For example, the existence of supersymmetric Schr\"odinger solutions is certainly of interest from the point of view of finding and classifying explicit supersymmetric solutions of supergravity and string theory. Furthermore, from a holographic perspective, the Schr\"odinger solutions provide excellent examples to study holography for non-asymptotically AdS spacetimes. Holographically they appear to be dual to very specific irrelevant deformations of conformal field theories which break Lorentz invariance but retain exact non-relativistic scale invariance \cite{Guica:2010sw,Costa:2010cn,balt}. In certain examples with $n=2$ these deformations should be dual to non-commutative `dipole' theories \cite{Adams:2008wt,Herzog:2008wg,Maldacena:2008wh}  but in many other cases the precise structure of the dual field theory is unknown.

Given the large class of Schr\"odinger deformations for the near horizon $AdS_5$ geometry of D3 branes it is natural to ask whether one can find similar solutions for the near horizon limit of the D1-D5 (or F1-NS5) system in type IIB supergravity. In this paper we address this question and using an Ansatz similar to the one in \cite{Bobev:2009mw} we find an infinite class of such solutions with integer and half integer dynamical exponents. The solutions are determined by a harmonic one-form and a scalar function on the $S^3$, and for specific choices of these we were able to construct solutions which preserve four supersymmetries. We will show that all our solutions with $n>1$ may be interpreted to leading order as corresponding to spin one and/or spin two irrelevant deformations of the holographically dual CFT.

The action of the Schr\"odinger algebra on these spacetimes is such that a possible dual condensed matter system has to live in \emph{two} dimensions less than the number of (asymptotically large) dimensions of the dual gravity solution \cite{Son:2008ye,Balasubramanian:2008dm}. In our case the non-compact part of the spacetime is always a deformation of AdS$_3$ and therefore according to the arguments of \cite{Son:2008ye,Balasubramanian:2008dm} it describes a one-dimensional (\emph{i.e.} quantum-mechanical) Schr\"odinger invariant system. This is a degenerate case and therefore the direct applicability of our solutions to quantum critical points is unclear. However one may alternatively view the spacetimes as corresponding to deformations of the two-dimensional D1-D5 (or F1-NS5) CFT dual to the undeformed $AdS_3$ solution. In this light our solutions are certainly not degenerate and they provide an interesting setting where the general questions we mentioned above may be studied in more detail.

The organization of the paper is as follows. In Section 2 we describe our Ansatz for the Schr\"odinger deformed backgrounds of IIB supergravity and show that the corresponding equations of motion reduce to a Laplace eigenvalue equation for a transverse vector harmonic on $S^3$ and a non-homogeneous scalar Laplace equation for the metric deformation. Then we proceed in Section 3 to present a number of explicit solutions and outline how to construct solutions for general dynamical exponent by using harmonic analysis on $S^3$. In Section 4 we study the supersymmetry variations for the dilatino and gravitino, derive the conditions for existence of supersymmetry and present some examples of supersymmetric solutions with 1/4 of the 16 supersymmetries preserved by $AdS_3\times S^3\times T^4$. In Section 5 we discuss our solutions from the point of view of the dual field theory. We present our conclusions and discuss some directions for further study in Section 6. In the Appendix we present the explicit Killing spinors on the undeformed $AdS_3\times S^3\times T^4$ background.

\textbf{Note added:} While we were preparing the manuscript we became aware of \cite{Kraus:2011pf}, which has some overlap with our results.

\section{Schr\"odinger deformations of $AdS_3\times S^3$}

Our aim is to find solutions of IIB supergravity which are deformations of the familiar $AdS_3\times S^3 \times T^4$ solution and have non-relativistic scaling symmetry with general dynamical exponent. To this end we consider the following IIB Ansatz (which is quite similar to the one studied in \cite{Bobev:2009mw}) 
\begin{eqnarray}
&& ds^2_{10} = R^2 \left(-\ds\frac{\Omega}{r^{2n}} du^2 + \ds\frac{-2dudv + dr^2}{r^2} + d\theta^2 + \sin^2\theta d\phi_1^2 + \cos^2\theta d\phi_2^2 \right) + ds^2_{T^4}~, \label{Ansatz1}\\
&& F^{(3)} =c_1 2 R^2 \left(\ds\frac{du\wedge dv \wedge dr}{r^3}  +  \sin\theta\cos\theta d\theta \wedge d\phi_1 \wedge d\phi_2\right)+ c_2d\left(\ds\frac{R^2}{r^n} \mathcal{A} \wedge du\right)~, \label{Ansatz2}\\
&& H^{(3)} = c_3 2 R^2\left(\ds\frac{du\wedge dv \wedge dr}{r^3}+  \sin\theta\cos\theta d\theta \wedge d\phi_1 \wedge d\phi_2 \right) + c_4d\left(\ds\frac{R^2}{r^n} \mathcal{A} \wedge du\right)~, \label{Ansatz3}
\end{eqnarray}
where $F^{(3)} =dC^{(2)}$ and $H^{(3)} =dB^{(2)}$ are the RR and NS-NS flux, $\Omega(\theta,\phi_1,\phi_2)$ is a function on $S^3$, $\mathcal{A}$ is a 1-form on $S^3$ and $ds^2_{T^4}$ is the  $T^4$ metric\footnote{Equivalently we can choose $K3$ instead of $T^4$ as internal manifold. Since $K3$ is hyper-K\"ahler the equations of motion and the supersymmetry analysis presented below will still be valid.}. The dynamical exponent is $n$ and as discussed below it is determined by the spectrum of the Laplacian on $S^3$ acting on scalars or vectors.

This Ansatz captures a number of familiar solutions. If we set $\mathcal{A}=\Omega=0$ and $c_3=0$ we recover the near horizon limit of a bound state of intersecting D1 and D5 branes which is a 1/2 BPS solution, \emph{i.e.} it preserves 16 supersymmetries. The metric is simply $AdS_3\times S^3\times T^4$ where the metric on $AdS_3$ is written in light-cone coordinates. This background has a holographic dual description in terms of the D1-D5 CFT, see \cite{David:2002wn} for a review. Similarly if we set $\mathcal{A}=\Omega=c_1=0$ we have the near horizon limit of intersecting F1 strings and NS5 branes. The two solutions are related by the $SL(2,R)$ symmetry of IIB supergravity, \cite{Schwarz:1983qr}, which is broken to $SL(2,Z)$ in string theory. The F1-NS5 solution has only NS-NS fields turned on and has a well known world-sheet description in terms of a $SL(2,R)^2\times SU(2)^2\times U(1)^4$ WZW model (see \cite{Orlando:2006cc} for a review).

We are interested in deforming these familiar backgrounds by switching on non-trivial flux and metric deformations determined by $\mathcal{A}$ and $\Omega$. To find such solutions we plug the above Ansatz in the equations of motion and Bianchi identities of IIB supergravity with vanishing dilaton-axion, which are given by \cite{Schwarz:1983qr}
\begin{eqnarray}
&& R_{ab} = \ds\frac{1}{4}\left( H^{(3)}_{acd} H_{b}^{(3)cd} + F^{(3)}_{acd} F_{b}^{(3)cd} -\ds\frac{g_{ab}}{12}  (H^{(3)}_{mnp} H^{(3)mnp} + F^{(3)}_{mnp} F^{(3)mnp} )\right) +\ds\frac{1}{6} F^{(5)}_{acdef}F_{b}^{(5)cdef}~, \notag \\\notag\\
&& d\star_{10} H^{(3)} = -4 F^{(5)} \wedge F^{(3)}~, \qquad d\star_{10} F^{(3)} = 4 F^{(5)} \wedge H^{(3)}~, \qquad dF^{(3)} = dH^{(3)} = 0~,\notag\\\label{IIBEoM}\\
&& d F^{(5)} = d\star_{10} F^{(5)} = - \ds\frac{1}{4} F^{(3)} \wedge H^{(3)}~, ~~~ H^{(3)}_{mnp} H^{(3)mnp} - F^{(3)}_{mnp} F^{(3)mnp}=0 = H^{(3)}_{mnp} F^{(3)mnp} ~, \notag\\\notag\\
&& F^{(5)} = \star_{10} F^{(5)} = dC^{(4)} - \ds\frac{1}{8} (C^{(2)} \wedge  H^{(3)} - B^{(2)} \wedge F^{(3)})~. \notag
\end{eqnarray}

The equations of motion restrict the constants $c_i$ as follows
\begin{equation}
c_1^2 + c_3^2 = 1~, \qquad\qquad c_2^2 +c_4^2 = 1~, \qquad\qquad c_1c_2+c_3c_4=0~, \label{c1234const}
\end{equation}
where we have assumed that either $c_2$ or $c_4$ are not zero. If $c_2=c_4=0$ the equations of motion require that $c_1^2 + c_3^2 = 1$.

For general $\mathcal{A}$ one finds that the Einstein equations for our Ansatz reduce to a single differential equation\footnote{We will adopt the notation $\nabla^2_{S^3}\equiv g^{\alpha\beta}\nabla_{\alpha}\nabla_{\beta}$, where $g_{\alpha\beta}$ is given by \eqref{unitS3met}.} 
\begin{equation}
\nabla^2_{S^3} \Omega + 4n(n-1) \Omega =  \ds\frac{1}{2}\mathcal{F}_{\alpha\beta} \mathcal{F}^{\alpha\beta} + n^2 \mathcal{A}_{\alpha} \mathcal{A}^{\alpha}~. \label{Omeqn}
\end{equation}
where $\mathcal{F}_{\alpha\beta} = \nabla_{\alpha}\mathcal{A}_{\beta} - \nabla_{\beta}\mathcal{A}_{\alpha}$ and the contraction over $\alpha$ and $\beta$ is performed with the metric $g_{\alpha\beta}$ on the unit radius $S^3$
\begin{equation}
g_{\alpha\beta}dx^{\alpha}dx^{\beta} = d\theta^2 + \sin^2\theta d\phi_1^2 + \cos^2\theta d\phi_2^2~.\label{unitS3met}
\end{equation}
The Maxwell equations reduce to 
\begin{eqnarray}
&& \nabla^{\alpha}\mathcal{A}_{\alpha} = 0~, \label{divA}\\
&& \nabla^{\alpha}\mathcal{F}_{\alpha\beta} + n^2 \mathcal{A}_{\beta} = 0 \label{divF}~.
\end{eqnarray}
One can use \eqref{divA} and the fact that the unit radius $S^3$ is an Einstein manifold, with curvature $R^{S^3}_{\alpha\beta} = 2 g_{\alpha\beta}$, to rewrite \eqref{divF} as
\begin{equation}
\nabla^2_{S^3} \mathcal{A}_{\alpha}  = (2-n^2) \mathcal{A}_{\alpha} ~. \label{curlyAeqn}
\end{equation}
So $ \mathcal{A}$ has to be a transverse (\emph{i.e.} $\nabla^{\alpha}\mathcal{A}_{\alpha} = 0$) vector spherical harmonic on $S^3$. The dynamical exponent of the solution is then determined by the spectrum of transverse vector spherical harmonics on $S^3$. We will discuss this further in Section 3. 

When $\mathcal{A}=0$ the IIB field equations reduce to the following equation for the function $\Omega$
\begin{equation}
\nabla^2_{S^3} \Omega = - 4n(n-1) \Omega~. \label{eqnOmA0}
\end{equation}
The deformed solution in this case has no flux deformation in \eqref{Ansatz2} and is determined by a scalar spherical harmonic on $S^3$. The allowed values for the dynamical exponent are determined by the spectrum of scalar harmonics on $S^3$.

To summarize, the IIB equations of motion \eqref{IIBEoM} for the Ansatz \eqref{Ansatz1}-\eqref{Ansatz3} reduce to two simple differential equations, \eqref{curlyAeqn} and \eqref{Omeqn}, on $S^3$ for the one-form $\mathcal{A}$ and the metric function $\Omega$, respectively. In addition we have to impose a transversality condition on $\mathcal{A}$, \eqref{divA}, and the constraints \eqref{c1234const} on the constants $\{c_1,c_2,c_3,c_4\}$.  In the next section we will discuss how to solve these equations and present a number of explicit solutions.

\section{Explicit solutions}

As we discussed in Section 2 the equations of motion of IIB supergravity for our Ansatz reduce to solving nonhomogeneous Laplace equations on $S^3$, so before we discuss explicit solutions let us start with a summary\footnote{ More details can be found in \cite{Deger:1998nm}.} on scalar and transverse vector spherical harmonics on $S^3$. 

The standard embedding coordinates of the unit radius $S^3$ in $\mathbb{R}^4$ are
\begin{equation}
Y^{1} = \sin\theta\cos\phi_1~, \qquad Y^{2} = \sin\theta\sin\phi_1~,\qquad Y^{3} = \cos\theta\cos\phi_2~, \qquad Y^{4} = \cos\theta\sin\phi_2~. \label{YAdef}
\end{equation}
These are in fact also the scalar spherical harmonics with eigenvalue $-3$ (see \eqref{scaharmspectrum} below). They satisfy the following useful identities
\begin{equation}
\nabla_{\alpha}Y^{A}_{\beta} = - g_{\alpha\beta} Y^{A}~, \qquad\qquad \ds\sum_{A=1}^{4}(Y^{A})^2=1~, \qquad\qquad g^{\alpha\beta}Y^{A}_{\alpha}Y^{B}_{\beta} = \delta^{AB}-Y^{A}Y^{B}~, \label{harmident}
\end{equation}
where we have defined $Y^{A}_{\alpha} \equiv \partial_{\alpha}Y^{A}$. Note that while $\nabla^2_{S^3}Y^{A}_{\alpha}=- Y^{A}_{\alpha}$ and therefore $Y^{A}_{\alpha}$ is a vector harmonic of eigenvalue $-1$, it is not transverse since $\nabla^{\alpha}Y^{A}_{\alpha} = - 3 Y^{A} \neq 0$. It is clear from \eqref{vecharmspectrum} below that the lowest transverse vector harmonic has eigenvalue $-2$.  We can build all scalar spherical harmonics and transverse vector spherical harmonics by taking products (and appropriate symmetrizations) of $Y^{A}$ and $Y^{A}_{\alpha}$. The scalar harmonic are given by the symmetric traceless products
\begin{equation}
Y^{A_1\ldots A_l} = Y^{(A_1}\ldots Y^{A_l)}~, \qquad\qquad l=0,1,2,\ldots \label{scalarharmsym}
\end{equation}
The transverse vector spherical harmonics are given by the following product
\begin{equation}
Y^{A_1\ldots A_{p}}_{\alpha} = Y^{[A_1}_{\alpha}Y^{(A_2]}\ldots Y^{A_p)}~, \qquad\qquad p=2,3,4,\ldots~, \label{vectorharmsym}
\end{equation}
where the indices are symmetrized according to the $SO(4)$ Young tableau with $p-1$ boxes in the first row and one box in the second. The spectrum of the Laplacian on these harmonics is then
\begin{eqnarray}
\nabla^2_{S^3}Y^{A_1\ldots A_l} &=& -l(l+2) Y^{A_1\ldots A_l}~,\qquad \qquad  l =0,1,2,\ldots~, \label{scaharmspectrum} \\
\nabla^2_{S^3}Y^{A_1\ldots A_{p}}_{\alpha} &=& - (p^2-2) Y^{A_1\ldots A_{p}}_{\alpha}~, \qquad\qquad p=2,3,4,\ldots~. \label{vecharmspectrum}
\end{eqnarray}
The degeneracy of the degree $l$ scalar harmonic is $(l+1)^2$ and the degeneracy of the degree $p$ transverse vector harmonic is $2(p^2-1)$. Comparing the equation \eqref{curlyAeqn} for $\mathcal{A}$  with \eqref{vecharmspectrum} we conclude that the spectrum of allowed dynamical exponents for the solutions with $\mathcal{A}\neq0$ is\footnote{We restrict our discussion to solutions with non-negative dynamical exponent.}
\begin{equation}
n=p~, \qquad\qquad p=2,3,4\ldots~. 
\end{equation}
In other words we have an infinite class of backgrounds given by \eqref{Ansatz1}-\eqref{Ansatz3} which are Schr\"odinger invariant with an integer dynamical exponent greater than or equal to two.

When $\mathcal{A}=0$ we only have to solve the harmonic equation \eqref{eqnOmA0} for $\Omega$, which upon comparison with \eqref{scaharmspectrum}, yields the following values of the dynamical exponent
\begin{equation}
n = \ds\frac{l}{2}+1~, \qquad\qquad l=0,1,2,\ldots~.
\end{equation}
Therefore we have a family of Schr\"odinger invariant solutions with half-integer and integer exponents and $n\geq1$. This should be contrasted with the IIB supergravity solutions discussed in \cite{Hartnoll:2008rs,Bobev:2009mw,Donos:2009xc} for which, even for $\mathcal{A}=0$, the dynamical exponent is an integer. There is also a special solution for which $\Omega$ is a constant and $n=0$. For this solution the three-dimensional metric in \eqref{Ansatz1} satisfies the vacuum Einstein equations with a negative cosmological constant and therefore it should be diffeomorphic to a part of $AdS_3$ since in three-dimensional gravity there are no propagating degrees of freedom. Solutions with $n=0$ in other dimensions were discussed in more detail in \cite{Costa:2010cn}.

It is curious to note that for any of the allowed values of $n$ there is a zero mode in the equation \eqref{Omeqn} for $\Omega$, \emph{i.e.} one is free to add to $\Omega$ a function $\Omega_{\text h}$ satisfying
\begin{equation}
\nabla^2_{S^3} \Omega_{\text h} = - 4n(n-1) \Omega_{\text h}~,
\end{equation}
so $\Omega_{\text h}$ is a scalar spherical harmonic of degree $l=2n-2$. We will discuss this further below.

We will end the general discussion on spherical harmonics by defining the following quantity which appears as the source term in \eqref{Omeqn},
\begin{equation}
\mathcal S = \frac{1}{2} \mathcal F_{\alpha \beta} \mathcal F^{\alpha \beta} + n^2 \mathcal A_\alpha \mathcal A^\alpha\,.
\end{equation}
The object $\mathcal{S}$ is clearly a scalar on $S^3$ but it carries $SO(4)$ indices if we write $\mathcal A_\alpha$ as a harmonic of the form \eqref{vectorharmsym}. Explicitly we find that $\mathcal{S}$ is a linear combination of the following $SO(4)$ tensors:
\begin{equation}
\mathcal{S}^{A_1\ldots A_n B_1\ldots B_n} = 2g^{\alpha\gamma}g^{\beta\delta}\nabla_{[\alpha}Y_{\beta]}^{A_1\ldots A_n} \nabla_{[\gamma}Y_{\delta]}^{B_1\ldots B_n} + n^2 g^{\alpha\beta}Y^{A_1\ldots A_n}_{\alpha}Y^{B_1\ldots B_n}_{\beta}~.\label{curlySgeneral}
\end{equation}
This expression for $\mathcal S$ will be useful below. 

\subsection{Solutions with $n=2$}

In this section we will construct explicit solutions of the form \eqref{Ansatz1}-\eqref{Ansatz3} with $n=2$ by following the general discussion on spherical harmonics above. The transverse vector spherical harmonics of degree $n=2$ are dual to the six independent Killing vectors on $S^3$ and are given by
\begin{equation}
Y_{\alpha}^{AB} = Y^{B}\partial_{\alpha}Y^{A}- Y^{A}\partial_{\alpha}Y^{B}~. \label{Kill1forms}
\end{equation}
The most general deformation in the flux \eqref{Ansatz2} for $n=2$ is given by the following linear combination
\begin{equation}
\mathcal{A}_{\alpha}= \eta_1Y_{\alpha}^{12} + \eta_2Y_{\alpha}^{13} +\eta_3Y_{\alpha}^{14} +\eta_4Y_{\alpha}^{23} +\eta_5Y_{\alpha}^{24} +\eta_6Y_{\alpha}^{34} ~. \label{generaln2harmonic}
\end{equation}
The expression for the tensor $\mathcal{S}$ for $n=2$ is particularly simple
\begin{equation}
\mathcal{S}^{A_1A_2 B_1B_2} = 4(\delta^{A_1B_1}\delta^{A_2B_2}-\delta^{A_1B_2}\delta^{A_2B_1})=8\delta^{A_1[B_1}\delta^{B_2]A_2}~. \label{curlySn2}
\end{equation}
With this in hand one finds 
\begin{equation}
\ds\frac{1}{2}\mathcal{F}_{\alpha\beta} \mathcal{F}^{\alpha\beta} + n^2 \mathcal{A}_{\alpha} \mathcal{A}^{\alpha} = 4 \ds\sum_{i=1}^{6}\eta_i^2~.\label{Omsourcen2}
\end{equation}
Plugging \eqref{Omsourcen2} in \eqref{Omeqn} we find that the nonhomogeneous solution for $\Omega$ is
\begin{equation}
\Omega_{\text{nh}}= \ds\frac{1}{2} \ds\sum_{i=1}^{6}\eta_i^2~.
\end{equation}
To this solution one can of course add a homogeneous solution which is a linear combination of scalar harmonics of the form \eqref{scalarharmsym} with degree $l=2$. The general solution for $\Omega$ is then
\begin{equation}
\Omega = \Omega_{\text{nh}}+ \Omega_{\text{h}}= \ds\frac{1}{2} \ds\sum_{i=1}^{6}\eta_i^2 + \ds\sum_{A,B=1}^{4}\xi_{AB}Y^{AB}~, \label{Omn2gen}
\end{equation}
where $\xi_{AB}$ are arbitrary constants. 

Schr\"odinger invariant solutions with $n=2$ can also be constructed by using a solution generating technique called the null Melvin twist\footnote{The null Melvin twist was originally applied to generate Schr\"odinger invariant deformations of $AdS_5\times S^5$ but it can be applied in much the same way to $AdS_3\times S^3$ \cite{Mazzucato:2008tr}.} \cite{Adams:2008wt,Herzog:2008wg,Maldacena:2008wh} along any Killing vector on $S^3$ \cite{Bobev:2009zf}. Applying the null Melvin twist to the Killing vector dual to $\mathcal{A}_{\alpha}$ as written in \eqref{generaln2harmonic} one finds that 
\begin{eqnarray}
&&\Omega_{\text{NMT}} = g^{\alpha\beta}\mathcal{A}_{\alpha}\mathcal{A}_{\beta}= [\cos\phi_1 (\eta_2\cos\phi_2+\eta_3\sin\phi_2)+\sin\phi_1 (\eta_4\cos\phi_2+\eta_5\sin\phi_2)]^2 \notag\\ &&+ [\eta_6\cos\theta + \sin\theta\cos\phi_2 (\eta_3\cos\phi_1+\eta_5\sin\phi_1)-\sin\theta\sin\phi_2 (\eta_2\cos\phi_1+\eta_4\sin\phi_1)]^2 \label{OmNMT}\\ &&+ [\eta_6\sin\theta + \cos\theta\sin\phi_1 (\eta_2\cos\phi_2+\eta_3\sin\phi_2)-\cos\theta\cos\phi_1 (\eta_4\cos\phi_2+\eta_5\sin\phi_2)]^2~. \notag 
\end{eqnarray}
where we put a subscript NMT to indicate that this is the function $\Omega$ generated by the twist. The solution generated by the particular null Melvin twist with $\eta_1=\eta_6$ and $\eta_2=\eta_3=\eta_4=\eta_5=0$ was discussed in \cite{Mazzucato:2008tr} and its thermal generalization was recently studied in \cite{Banerjee:2011jb}.

At face value the two expressions for $\Omega$, \eqref{Omn2gen} and \eqref{OmNMT}, look quite different and this may appear puzzling since these solutions have the same $\mathcal{A}$. There is of course no contradiction between the two solutions with $n=2$.  To see this one realizes that the function 
\begin{equation}
\mathcal{H} = g^{\alpha\beta}\mathcal{A}_{\alpha}\mathcal{A}_{\beta} - \ds\frac{1}{2} \ds\sum_{i=1}^{6}\eta_i^2 ~,
\end{equation}
satisfies the equation
\begin{equation}
\nabla^2_{S^3}\mathcal{H} = -8 \mathcal{H}~,
\end{equation}
in other words it is a linear combination of scalar harmonics on $S^3$ of degree $l=2$. Therefore \eqref{Omn2gen} can be rewritten as
\begin{equation}
\Omega = \ds\frac{1}{2} \ds\sum_{i=1}^{6}\eta_i^2 + \gamma \mathcal{H} + \ds\sum_{A,B=1}^{4}\widetilde{\xi}_{AB}Y^{AB}~, \label{Omn2gennew}
\end{equation}
where $\widetilde{\xi}_{AB}$ are some new constants. Comparing \eqref{OmNMT} with \eqref{Omn2gennew} we see that the null Melvin twist has generated a particular solution with $\gamma=1$ and $\widetilde{\xi}_{AB}=0$ and there is indeed no contradiction between \eqref{Omn2gen} and \eqref{OmNMT}.

To the best of our knowledge there are no dualities in string theory that provide solution generating techniques for Schr\"odinger invariant backgrounds with $n\neq2$ and one will have to solve explicitly equations \eqref{Omeqn} and \eqref{curlyAeqn} in order to find solutions with $n>2$. This will be the subject of the next two sections.

\subsection{Solutions with $n=3$}

For $n=3$ the one-form $\mathcal{A}$ is a linear combination of the following transverse vector harmonics on $S^3$:
\begin{equation}
Y_{\alpha}^{ABC} = 2Y^{B}Y^{C}\partial_{\alpha}Y^{A}- Y^{A}Y^{C}\partial_{\alpha}Y^{B}- Y^{A}Y^{B}\partial_{\alpha}Y^{C}~. \label{n3vecharmonic}
\end{equation}
For a given linear combination of the harmonics \eqref{n3vecharmonic} the source term in the equation for $\Omega$ is a linear combination of the following tensors
\begin{multline}
\mathcal{S}^{A_1A_2 A_3 B_1B_2B_3} = 18( \delta^{A_1[B_1}\delta^{B_3]A_3} Y^{A_2}Y^{B_2} + \delta^{A_1[B_1}\delta^{B_2]A_2} Y^{A_3}Y^{B_3} + \delta^{A_1[B_1}\delta^{B_2]A_3} Y^{A_2}Y^{B_3} \\+ \delta^{A_1[B_1}\delta^{B_3]A_2} Y^{A_3}Y^{B_2})~. \label{curlySn3}
\end{multline}
In the simple case in which we choose $\mathcal{A}_{\alpha}=Y_{\alpha}^{ABC}$ (for a given choice of the indices $A,B,C$) the nonhomogenous solution for $\Omega$ is
\begin{equation}
\Omega_{\text{nh}} = \ds\frac{9}{16} \left[(Y^{B})^2+(Y^{B})^2 - \ds\frac{1}{6}\right]~. \label{omegan3squares}
\end{equation}
There are 16 independent vector harmonics of the form \eqref{n3vecharmonic} and, using  \eqref{curlySn3}, one can show that there is no linear combination for which $\Omega_{\text{nh}}$ is a constant. It is possible to express \eqref{omegan3squares} in terms of scalar harmonics on $S^3$. Using \eqref{harmident} and \eqref{scalarharmsym} it is not hard to rewrite \eqref{omegan3squares} as
\begin{equation}
\Omega_{\text{nh}} = \ds\frac{9}{16} \left[Y^{BB}+Y^{CC} +\ds\frac{1}{3}\right]~. \label{omegan3harmonics}
\end{equation}
Therefore the nonhomogeneous solution for $\Omega$ is a linear combination of two scalar harmonics with $l=2$ and one with $l=0$. As we already emphasized above, one can add a homogenous scalar harmonic with eigenvalue $-24$, \emph{i.e.} $l=4$, to $\Omega$. 

It is clear from \eqref{YAdef} and \eqref{omegan3squares} that $\Omega$ can change sign in some domain on $S^3$. The same behavior of $\Omega$ was noted also for some of the solutions  in \cite{Hartnoll:2008rs,Bobev:2009mw}. We refer the reader to \cite{Hartnoll:2008rs} for a more detailed discussion on the consequences of this fact.

\subsection{General dynamical exponents}

After having presented two explicit examples we would like to outline the general procedure for solving the nonhomogeneous equation for $\Omega$. First one chooses a transverse vector spherical harmonic, $\mathcal{A}$, of degree $p$ which would be a linear combination of the vector spherical harmonics $Y_{\alpha}^{A_1\ldots A_p}$. This choice determines the dynamical exponent of the solution to be $n=p$. The next step is to compute the source term $\mathcal{S}=\frac{1}{2}\mathcal{F}^2 + n^2\mathcal{A}^2$ by taking an appropriate linear combination of terms like \eqref{curlySgeneral}. Since $\mathcal{A}$ and $\mathcal{F}$ are of degree $p$ the naive conclusion is that $\mathcal{S}$ is a polynomial of degree $2p$ in the elementary spherical harmonics $Y^{A}$. However using the identities for harmonic functions \eqref{harmident}, \eqref{scalarharmsym} and \eqref{vectorharmsym} one can show that $\mathcal{S}$ is in fact a polynomial of degree $2p-4$ in the $Y^{A}$'s with only even powers. This is also illustrated by the expressions for $\mathcal{S}$ in the examples with $n=2$ and $n=3$ studied above which are equations \eqref{curlySn2} and \eqref{curlySn3}, respectively. To solve for $\Omega$ we should therefore expand it as a linear combination of scalar spherical harmonics up to order $2p-4$ and solve for the coefficients. In this way the problem of finding an explicit solution becomes a problem in linear algebra. It is important to emphasize that the source in the equation for $\Omega$ is of degree $2p-4$ whereas the homogeneous solution of the same equation is of degree $2p-2$. This proves that for a given choice of $\mathcal{A}$ we will always be able to find a nonhomogeneous solution for $\Omega$ of degree $\leq 2p-4$ in addition to the homogeneous solution.

It will be interesting to understand whether there are other values of $n$, besides $n=2$, for which there are solutions with constant $\Omega$. Upon dimensional reduction on $S^3\times T^4$ such solutions should yield three-dimensional truncated models with massive vector fields in which the mass of the vector field is related to $n$. Using \eqref{curlySn3} we have showed that there is no linear combination of vector harmonics which yields constant $\Omega$ for $n=3$ and we were not able to find such a linear combination for $n=4$. This suggests that the case $n=2$ might be special and there are no solutions with constant $\Omega$ for $n>2$.

\section{Supersymmetry analysis}

As discussed in \cite{Hartnoll:2008rs,Donos:2009en,Bobev:2009mw,Donos:2009xc,Donos:2009zf} there are many supersymmetric Schr\"odinger invariant deformations of the Freund-Rubin $AdS_5$ compactifications of IIB supergravity and it is natural to ask whether the solutions presented in the previous section preserve some of the 16 supersymmetries of the undeformed $AdS_3\times S^3 \times T^4$. In this section we will address this question and will indeed find that a number of our solutions preserve 4 supersymmetries.

We will use the supersymmetry variations of IIB supergravity as written in \cite{Gauntlett:2005ww} and will choose a basis with purely real gamma matrices. The dilatino and gravitino variations for the background \eqref{Ansatz1}-\eqref{Ansatz3} are
\begin{eqnarray}
\delta\lambda &=& \ds\frac{i}{24} \Gamma^{MNP}(H^{(3)}_{MNP}-i F^{(3)}_{MNP}) \epsilon^{*}= 0~,\\
\delta\psi_{\mu} &=& \nabla_{\mu}\epsilon + \ds\frac{1}{96} \left[ \Gamma_{\mu}^{MNP}(H^{(3)}_{MNP} + i F^{(3)}_{MNP}) -9 \Gamma^{MN}(H^{(3)}_{\mu MN} + i F^{(3)}_{\mu MN})\right]\epsilon^{*} = 0~,
\end{eqnarray}
where $\epsilon$ is a complex 32-component chiral spinor,
\begin{equation}
(1+ \Gamma^{11})\epsilon=0~, \label{10Dchirality}
\end{equation}
$\epsilon^{*}$ is the complex conjugate of $\epsilon$ and we are using the following standard notation
\begin{equation}
\nabla_{\mu}\epsilon = \partial_{\mu}\epsilon + \ds\frac{1}{4}\omega_{\mu MN}\Gamma^{MN}\epsilon ~, \qquad \Gamma^{M_1 \ldots M_k} = \Gamma^{[M_1}\ldots \Gamma^{M_k]}~, \qquad \Gamma^{11} \equiv \Gamma^{123\ldots 10}~.
\end{equation}
The gamma matrices satisfy the algebra
\begin{equation}
\{\Gamma^{\mu},\Gamma^{\nu}\} = 2 G^{\mu\nu}~,
\end{equation}
where $G_{\mu\nu}$ is the ten-dimensional metric in \eqref{Ansatz1}. Our vielbeins are\footnote{In the remainder of this paper, for convenience, we choose to work with unit radius $AdS_3$ and $S^3$, \emph{i.e.} we set $R=1$ in \eqref{Ansatz1}.}
\begin{eqnarray}
e^1 &=& \ds\frac{\Omega+1}{2r^n}du + \ds\frac{1}{r^{2-n}}dv~, \qquad e^2 = \ds\frac{\Omega-1}{2r^n}du + \ds\frac{1}{r^{2-n}}dv~, \qquad e^3 = \ds\frac{dr}{r}~,\\
e^4 &=& d\theta~, \qquad e^5 = \sin\theta d\phi_1~, \qquad e^6 = \cos\theta d\phi_2~, \qquad e^{6+a} = dy^a~, ~~~ a=1,2,3,4~.
\end{eqnarray}
The dilatino variation is
\begin{equation}
\ds\frac{i(c_3-ic_1)}{2}(\Gamma^{123}+ \Gamma^{456})\epsilon^{*} + \ds\frac{i(c_4-ic_2)}{8} \mathcal{M} (\Gamma^{1}- \Gamma^{2})\epsilon^{*} =0~, \label{dilatinovar}
\end{equation}
where we have defined the matrix
\begin{equation}
\mathcal{M} \equiv \mathcal{F}_{MN}\Gamma^{MN} + 2n \mathcal{A}_{N}\Gamma^{N3}~. \label{curlyM}
\end{equation}
We can solve the dilatino variation by imposing that $\epsilon$ satisfies:
\begin{equation}
(1+ \Gamma^{123456})\epsilon=0~, \qquad\qquad \mathcal{M} (\Gamma^{1}- \Gamma^{2})\epsilon=0~. \label{dilatinoprojectors}
\end{equation}
The gravitino variations yield\footnote{Note that by virtue of the second constraint in \eqref{dilatinoprojectors} the last term in equations \eqref{gravth}-\eqref{gravph2} vanishes but we presented it for completeness.}
\begin{eqnarray}
\delta\psi_{y_a}&=&\partial_{y_a}\epsilon =0~, \qquad\qquad a=1,2,3,4~,\label{gravya}\\
\delta\psi_{\theta}&=&\partial_{\theta}\epsilon - \ds\frac{c_3+ic_1}{2} \Gamma^{56}\epsilon^{*}  \notag\\&&+ \ds\frac{c_4+ic_2}{4}\,\Gamma^4\left[\mathcal{F}_{56}\Gamma^{56}+n(\mathcal{A}_{5}\Gamma^{53}+\mathcal{A}_{6}\Gamma^{63}) -\ds\frac{3}{8}\mathcal{M}\right](\Gamma^1-\Gamma^2)\epsilon^{*}=0~,\label{gravth}\\
\delta\psi_{\phi_1}&=&\partial_{\phi_1}\epsilon - \ds\frac{\cos\theta}{2}\,\Gamma^{45}\epsilon + \ds\frac{c_3+ic_1}{2} \sin\theta\,\Gamma^{46}\epsilon^{*}  \notag\\&&+ \ds\frac{c_4+ic_2}{4}\, \sin\theta\, \Gamma^5\left[\mathcal{F}_{46}\Gamma^{46}+n(\mathcal{A}_{4}\Gamma^{43}+\mathcal{A}_{6}\Gamma^{63}) -\ds\frac{3}{8}\mathcal{M}\right](\Gamma^1-\Gamma^2)\epsilon^{*}=0~,\label{gravph1}\\
\delta\psi_{\phi_2}&=&\partial_{\phi_2}\epsilon+ \ds\frac{\sin\theta}{2}\,\Gamma^{46}\epsilon - \ds\frac{c_3+ic_1}{2} \cos\theta\,\Gamma^{45}\epsilon^{*} \notag\\&&+ \ds\frac{c_4+ic_2}{4} \cos\theta\, \Gamma^6\left[\mathcal{F}_{45}\Gamma^{45}+n(\mathcal{A}_{4}\Gamma^{43}+\mathcal{A}_{5}\Gamma^{53}) -\ds\frac{3}{8}\mathcal{M}\right](\Gamma^1-\Gamma^2)\epsilon^{*} =0~,\label{gravph2}\\
\delta\psi_{r}&=&\partial_{r}\epsilon + \ds\frac{n-1}{2r}\,\Gamma^{12}\epsilon - \ds\frac{c_3+ic_1}{2r}\, \Gamma^{12}\epsilon^{*} \notag\\&&+ \ds\frac{c_4+ic_2}{32r} \left[\Gamma^{3}\mathcal{M}(\Gamma^{1}-\Gamma^{2})+8n\mathcal{A}_{N}\Gamma^{N}(\Gamma^1-\Gamma^2)\right]\epsilon^{*} =0~,\label{gravr}\\
\delta\psi_{v}&=&\partial_{v}\epsilon + \ds\frac{1}{2r^{2-n}}(\Gamma^{1}-\Gamma^{2})\Gamma^{3}\epsilon + \ds\frac{(c_3+ic_1)}{2r^{2-n}}(\Gamma^{1}-\Gamma^{2})\Gamma^{3}\epsilon^{*}=0~.\label{gravv}\\
\delta\psi_{u}&=&\partial_{u}\epsilon + \ds\frac{1}{4r^{n}}(\Gamma^{1}+\Gamma^{2})\Gamma^{3}\epsilon - \ds\frac{c_3+ic_1}{4r^{n}}(\Gamma^{1}+\Gamma^{2})\Gamma^{3}\epsilon^{*} \notag\\ &&+\ds\frac{(2n-1)\Omega}{4r^{n}}(\Gamma^{1}-\Gamma^{2})\Gamma^{3}\epsilon + \ds\frac{(c_3+ic_1)\Omega}{4r^{n}}(\Gamma^{1}-\Gamma^{2})\Gamma^{3}\epsilon^{*} \label{gravu}\\
&&-\ds\frac{1}{4r^n}\left[\partial_{\theta}\Omega\Gamma^{4}+\ds\frac{\partial_{\phi_1}\Omega}{\sin\theta}\Gamma^{5}+\ds\frac{\partial_{\phi_2}\Omega}{\cos\theta}\Gamma^{6}\right](\Gamma^{1}-\Gamma^{2})\epsilon-\ds\frac{c_4+ic_2}{32r^n}(3-\Gamma^{12})\mathcal{M}\epsilon^{*}=0~.\notag
\end{eqnarray}
Solving these equations in complete generality is a daunting task.\footnote{ For $\mathcal A = \Omega = 0$ these equations reduce to those for $AdS_3 \times S^3 \times T^4$. In the Appendix we give an explicit solution for that case and obtain the expected 16 supersymmetries.} However a fairly complete supersymmetry analysis can be performed if we solve the dilatino variation by imposing the first constraint in \eqref{dilatinoprojectors} and then solve the second constraint in \eqref{dilatinoprojectors} by requiring that
\begin{equation}
(\Gamma^{1}- \Gamma^{2})\epsilon=0~. \label{G12projector}
\end{equation}
With this simplifying assumption we can solve the first nine gravitino variations \eqref{gravya}-\eqref{gravv} explicitly. To this end we define
\begin{equation}
\epsilon = e^{i\chi}(\epsilon_1+ i\epsilon_2)~, \label{epse1e2}
\end{equation}
where using \eqref{c1234const} we have defined
\begin{equation}
c_3+ic_1 = e^{i2\chi}~, \qquad\qquad \chi\in\mathbb{R}~.
\end{equation}
An explicit solution to \eqref{gravya}-\eqref{gravv} is then given by
\begin{eqnarray}
\epsilon_1 &=&  e^{\frac{\theta}{2}\Gamma^{56}} e^{\frac{\phi_1 + \phi_2}{2}\Gamma^{45}} r^{\frac{2-n}{2}\Gamma^{12}}\zeta_1~, \qquad \qquad \epsilon_2 = 0~,
\end{eqnarray}
where $\zeta_1$ is a constant real spinor satisfying the ten-dimensional chirality projector \eqref{10Dchirality}, the first projector in \eqref{dilatinoprojectors} and the projector \eqref{G12projector} and therefore it has four independent components. The spinor $\epsilon_2$ has to vanish due to the projector \eqref{G12projector} and the constraints from the gravitino variation \eqref{gravu}.

To solve the remaining gravitino variation \eqref{gravu} we have to impose the additional condition
\begin{equation}
\mathcal{M}\epsilon=0~, \label{curlyMproj}
\end{equation}
where we recall that $\mathcal M$ was defined in \eqref{curlyM}. Using the first projector in \eqref{dilatinoprojectors} and \eqref{G12projector} it is not hard to show that
\begin{equation}
\mathcal{M}\epsilon = 2[(\mathcal{F}_{45}+n \mathcal{A}_{6})\Gamma^{45} + (\mathcal{F}_{46}-n \mathcal{A}_{5})\Gamma^{46} + (\mathcal{F}_{56}+n \mathcal{A}_{4})\Gamma^{56}]\epsilon~. \label{simplecurlyM}
\end{equation}
For a generic choice of $\mathcal{A}$ the matrix $\mathcal{M}$ will have non-vanishing determinant, equation \eqref{curlyMproj} will have no nontrivial solutions and there will be no supersymmetry preserved by the corresponding supergravity background. We therefore conclude that the condition \eqref{curlyMproj} has to be studied case by case for particular solutions. For the different solutions studied in Section 3 we obtained the following results:

\begin{itemize}

\item 

It is clear that for the supergravity solutions with $\mathcal{A}=0$ and any choice of the dynamical exponent $n$ the matrix $\mathcal{M}$ vanishes and the condition \eqref{curlyMproj} is trivially satisfied. This infinite class of solutions therefore preserves four supersymmetries, \emph{i.e.} 1/4 of the supersymmetries preserved by $AdS_3\times S^3\times T^4$.

\item 

For the solutions with $n=2$ and $\mathcal{A}$ given by \eqref{generaln2harmonic} the determinant of $\mathcal{M}$ vanishes only when all three of the following conditions are satisfied
\begin{equation}
\eta_6=-\eta_1~, \qquad \eta_5=\eta_2~,\qquad \eta_4=-\eta_3 ~. \label{susyetasn2}
\end{equation}
We then find, after using \eqref{simplecurlyM}, that for the choice of parameters \eqref{susyetasn2} the condition \eqref{curlyMproj} is trivially satisfied, therefore we have no additional constraints imposed on $\epsilon_1$ and the supergravity solution given by \eqref{generaln2harmonic}, \eqref{Omn2gen} and \eqref{susyetasn2} preserves four supersymmetries. As discussed in Section 3 we can add a scalar spherical harmonic $\Omega_{\text h}$ with eigenvalue $-4$ to $\Omega$ for $n=2$ and we still have a solution to the equations of motion. From the above discussion one finds that the addition of such a scalar harmonic does not break any additional supersymmetry.

\item 

We have analyzed in detail the solutions with non-zero $\mathcal{A}$ and $n=3$ and $n=4$. There are 16 independent transverse vector harmonics for $n=3$ and 30 for $n=4 $. We have showed that in both cases there is no linear combination of harmonics for which $\det(\mathcal{M})=0$. Therefore these solutions do not preserve any supersymmetry. Based on this analysis it is tempting to speculate that the solutions with $n>2$ and $\mathcal{A}\neq0$ do not have Killing spinors which satisfy the projection condition \eqref{G12projector}.

\end{itemize}

It is important to note that there are two caveats in our approach to solving the supersymmetry variations \eqref{dilatinovar} and \eqref{gravya}-\eqref{gravu}. First we have solved \eqref{dilatinovar} by imposing both projectors in \eqref{dilatinoprojectors}, but there is the logical possibility that for some solutions one can solve the dilatino variation by imposing a single projector. Although we believe imposing \eqref{dilatinoprojectors} is a necessary condition for solving the gravitino variations we have not been able to furnish a proof\footnote{Note that to solve the supersymmetry variations on the undeformed $AdS_3\times S^3\times T^4$ background we have to impose the six-dimensional chirality projector given by the first equation in \eqref{dilatinoprojectors}, see the Appendix for more details.}. The second assumption we have made is that the condition \eqref{G12projector} is satisfied. In principle there could be another class of solutions to the Killing spinor equations for which we solve the dilatino variation by imposing $\mathcal{M}\epsilon=0$ and requiring $(\Gamma^{1}-\Gamma^{2})\epsilon\neq0$. The analysis of the supersymmetry equations \eqref{gravya}-\eqref{gravu} in this case is much more involved and we cannot offer any general statements\footnote{Supersymmetric Schr\"odinger deformations of $AdS_5\times SE_5$ with this type of Killing spinors were studied in \cite{Donos:2009zf}.}. Therefore, although a preliminary non-exhaustive study for $n=2$ and $\mathcal{A}\neq0$ did not result in any Killing spinors that satisfy $(\Gamma^{1}-\Gamma^{2})\epsilon\neq0$, at this point we cannot exclude supersymmetric Schr\"odinger solutions with more than four supercharges.

\section{Comments on the dual CFT}

Solving the type IIB equations of motion for the Ansatz \eqref{Ansatz1}-\eqref{Ansatz3} reduces to solving a linear equation for $\mathcal A$ and subsequently a linear equation for $\Omega$ with a source quadratic in $\mathcal{A}$. This in particular implies that for every solution to the reduced equations \eqref{Omeqn} and \eqref{curlyAeqn} we can continuously rescale $\mathcal A$ and $\Omega$ to zero and recover the familiar $AdS_3 \times S^3 \times T^4$ solution. This spacetime has a well-defined CFT dual and switching on a small $\mathcal A$ and/or $\Omega$ can therefore be interpreted holographically by using the standard AdS/CFT dictionary. As was already observed in \cite{Son:2008ye} and subsequently worked out in more detail in \cite{Guica:2010sw,Costa:2010cn}, one finds that at least to lowest order the deformed spacetimes correspond to switching on a constant null source for an irrelevant spin one or spin two operator in the dual CFT. In the remainder of this section we will work out some of details of this deformation picture.

\subsection{The Kaluza-Klein reduction}

To determine more precisely which operators are sourced in the dual CFT we need to first Kaluza-Klein reduce the six-dimensional $\mathcal A$ and $\Omega$ over the $S^3$ to obtain three-dimensional bulk fields. The first step is to introduce the Kaluza-Klein decomposition, for example for $\Omega$ we would write:
\begin{equation}
\Omega = \sum_{l,m} \omega^{(l,m)} Y^{l,m}~,
\end{equation}
where $m$ labels the constituents of the $SO(4)$ multiplet of scalar $S^3$ harmonics $Y^{(l,m)}$ of degree\footnote{In this section we will use a shorthand index $(l,m)$, instead of $A_1 A_2 \ldots A_l$, to denote the scalar spherical harmonics \eqref{scalarharmsym} of degree $l$. The same will hold for the index $(n,q)$ of transverse vector harmonics in \eqref{curlyAexpand}. For a fixed $l$ the index $m$ goes from $1$ to $(l+1)^2$ and for a fixed $n$, $q$ goes from $1$ to $2(n^2-1)$.} $l$ and eigenvalue $-l(l+2)$. In particular, we use $Y^{0,1}$ to denote the constant function on the $S^3$. For our solutions the coefficients $\omega^{(l,m)}$ are all constants. For $\mathcal A$ we may use the fact that it is harmonic on the three-sphere with eigenvalue $2-n^2$, so its harmonic decomposition is given by:
\begin{equation}
\mathcal A_\alpha  = \sum_{m} b^{(n,q)} Y_\alpha^{n,q}~, \label{curlyAexpand}
\end{equation}
where now $q$ labels the $SO(4)$ multiplet of transverse vector harmonics $Y_\alpha^{n,q}$ with eigenvalue $2 - n^2$ and the coefficients $b^{(n,q)}$ are again constants.

Unfortunately, the coefficients $b^{(n,q)}$ and $\omega^{(l,m)}$ are not directly in one-to-one correspondence with components of three-dimensional supergravity fields. Instead, the map between the coefficients of the spherical harmonic decomposition of the six-dimensional (or more properly the ten-dimensional) fields and the eventual three-dimensional fields is well-known to be non-linear. As is shown for example in \cite{Skenderis:2006uy}, these nonlinearities are crucial in extracting the correct field theory data from a given supergravity solution. Without a detailed analysis of the non-linear Kaluza-Klein map we cannot therefore reliably say which three-dimensional supergravity fields are switched on and correspondingly which operators have a nonzero source and/or vev in the dual CFT.

Nevertheless we believe it is instructive to consider the Kaluza-Klein map in some more detail at the linearized level, even though we will not be able to draw any quantitative conclusions. The coefficients $b^{(n,q)}$ then correspond to a set of three-dimensional spin one fields which we call $\chi_\mu^{(n,q)}$, the coefficients $\omega^{(l,m)}$ with $l > 0$ correspond to massive spin two fields $\phi_{\mu \nu}^{(l,m)}$, and finally $\omega^{(0,1)}$ corresponds to a deformation of the three-dimensional AdS$_3$ metric $g_{\mu \nu}$. To first order the profile of these fields is given by:
\begin{equation}
\label{eq:3dfields}
\begin{split}
\chi_{\mu}^{(n,q)} &= \frac{b^{(n,q)}}{r^n} \delta_\mu^u + \ldots \\
\phi_{\mu \nu}^{(l,m)} &= - \frac{\omega^{(l,m)}}{r^{2n}} \delta_\mu^u \delta_\nu^u + \ldots \\
\delta g_{\mu \nu} &= - \frac{\omega^{(0,1)}}{r^{2n}} \delta_\mu^u \delta_\nu^u + \ldots
\end{split}
\end{equation}
For a given supergravity solution $n$ is fixed whereas $q$, $l$ and $m$ can take several values. The terms represented by the dots on the right-hand sides of \eqref{eq:3dfields} reflect our ignorance of the complete Kaluza-Klein map. We emphasize that for the spin two fields these terms are not necessarily subleading in the perturbation. (On the other hand, because of the lightlike nature of our deformation it may well be that the non-linear terms vanish beyond the first subleading order but this remains to be worked out.)

\subsection{The operator-field correspondence}

According to the usual AdS/CFT dictionary each of the three-dimensional bulk fields in \eqref{eq:3dfields} corresponds to a field theory operator with a specific scaling dimension $\Delta$. The value of $\Delta$ for the fields of \eqref{eq:3dfields} can be computed using the following procedure. The eigenvalue of the spherical harmonic (determined by $n$ or $l$) determines the mass of the three-dimensional field. This mass directly determines the allowed asymptotic radial behavior of the solution to the linearized field equations, from which we can in turn determine the scaling dimension of the dual operator.

As we mentioned above, for the massive vectors $\chi_\mu^{(n,q)}$ the corresponding eigenvalue is $2 - n^2$. We then find that from \cite{Deger:1998nm} that the three-dimensional mass $m_{v}^2 = n^2$. It is then not hard to solve the Proca equation for the massive vectors $\chi_\mu^{(n,q)}$ in an asymptotically AdS$_3$ spacetime and recover that the field behaves as $r^{\pm n}$ when we approach the boundary $r \to 0$. According to the usual AdS/CFT dictionary, the coefficient of the leading term $r^{-n}$ corresponds to the source for a vector operator. If a vector operator in a $d$-dimensional field theory (so in our case $d=2$) has scaling dimension $\Delta$, then its source with a lower vector index has dilatation weight $d - \Delta - 1$. This dilatation weight is equal to the power of $r$ and we therefore read off that
\begin{equation}
\Delta = 1 + n
\end{equation}
for the vector operators dual to $\chi_\mu^{(n,q)}$. Moreover, from \eqref{eq:3dfields} we find that the only nonzero term in the radial expansion of $\chi_\mu^{(n,q)}$ sits precisely at the order $r^{-n}$ corresponding to the source term. We therefore conclude that for nonzero $\mathcal A$ we have indeed deformed the dual CFT; more precisely, to first order we switched on sources $b^{(n,q)} \delta_i^u$ for vector operators $\mathcal O_i$ of scaling dimensions $\Delta = 1 +n$.

Let us now turn to the spin two fields. The fields $\phi_{\mu \nu}^{(l,m)}$, corresponding to a spherical scalar harmonic of eigenvalue $-l(l+2)$, have mass $m^2_{t} = l(l+2)$ \cite{Deger:1998nm}. The linear massive spin two equations in $AdS_{d+1}$ were solved for example in \cite{Polishchuk:1999nh}. For a massive spin two field with both indices down one finds asymptotic behavior of the form $r^{-\alpha}$ with
\begin{equation}
\alpha = 2 - \frac{d}{2} \pm \frac{1}{2} \sqrt{d^2 + 4 m_{t}^2}\,.
\end{equation}
In our case $d=2$ and $ m_{t}^2 = l(l+2)$ so the leading behavior becomes:
\begin{equation}
\label{eq:alphasource}
\alpha = l + 2~,
\end{equation}
and the subleading (normalizable) behavior becomes $\alpha = -l$ but it will not be important in what follows. With both indices down, the source for a spin two operator of dimension $\Delta$ has dilatation weight $d - \Delta - 2$ so we read off that:
\begin{equation}
\Delta = l + 2~,
\end{equation}
is the scaling dimension of the operator dual to $\phi_{\mu \nu}^{(l,m)}$.

Let us now return to the spin two part of the linearized version of the reduced solution \eqref{eq:3dfields}. Clearly according to \eqref{eq:alphasource} a term proportional to $r^{-2n}$ is only a source term when $2n = l + 2$. For all other values of $l$ this term is not a source term and neither is it, generically, a vev term. Instead it is merely an induced term in the radial expansion which is only present because of the vector deformation. Such a term therefore would \emph{not} correspond to switching on a source for the corresponding operator. Within our linearized analysis, then, only the fields $\phi_{\mu \nu}^{(2n-2,m)}$ correspond to nonzero sources $\omega^{(2n-2,m)}\delta_i^u \delta_j^u$ for spin two operators $\mathcal O_{ij}$ of scaling dimension $2n$. Notice that $2n-2$ is precisely the order of the scalar spherical harmonic $\Omega_{\text{h}}$ which is a homogeneous solution to the equation for $\Omega$. Therefore, for all $n$ the spin two deformation is independent of the spin one deformation and can be set to an arbitrary value by changing the coefficient of the homogenous solution for $\Omega_{\text h}$.

Combining now the spin one and the spin two results, we conclude that the solutions we have found can be interpreted holographically as deformations of the D1-D5 CFT by constant null sources given by $b^{(n,q)}$ for a class of vector operators as well as constant null sources given by $\omega^{(2n-2,m)}$ (which correspond to the homogeneous solution to $\Omega_{\text{h}}$), for a class of spin two operators.

Let us now comment on the aforementioned induced terms in the other fields and in particular in the metric, which at least in our linearized analysis do not correspond to field theory sources. Such induced terms are often present in AdS/CFT, for example they generically appear in the radial expansion of the metric when gravity is coupled to scalars. In the present cases, however, the power of $r$ is such that the induced term is \emph{more} leading than the source term for all $l < 2n -2$. In particular for $l=0$ we find that, for all $n > 1$, $\delta g_{\mu \nu}$ is more leading than the term corresponding to the boundary metric (\emph{i.e.} the usual part $-2 dudv/r^2$ which is also present in the AdS$_3$ metric). Notice that this phenomenon occurs precisely because the vector operator becomes irrelevant for $n > 1$; indeed, irrelevant deformations modify the UV behavior of the dual theory and correspondingly the bulk solution is no longer asymptotically of the AdS form. We believe that it is worthwhile to point out that these more leading induced terms still do not have to correspond to sources for any of the dual operators. Indeed, in the analysis of \cite{balt} such more leading terms were found at a perturbative level and yet the source of the dual operator was not modified in all the examples considered in that paper.

Let us end this section with a brief discussion of the linearized holographic analysis for the specific examples considered above. First of all, the (supersymmetric) solutions with $\mathcal A = 0$ and half-integer $n$ correspond to a deformation of the field theory with a spin two operator of scaling dimensions $\Delta=n+2$. (The cases $n=0$ and $n=1$ are special as they are diffeomorphic to a part of AdS$_3 \times S^3$.) Second, the solutions with $n=2$ and nonzero $\mathcal A$ correspond to a deformation with a massive vector operator of scaling dimension 3. We may pick this deformation such that $\Omega$ is constant and some supersymmetry is preserved and in that case we find that in the bulk no other spin two field besides the metric is switched on. We however also have the additional freedom to add a harmonic part to $\Omega$, \emph{i.e.} to switch on a constant source for massive spin two operators of dimension 4. This extra deformation preserves all the remaining supersymmetries. For $n=3$ we have no supersymmetries and find a deformation with a vector operator of dimension 4. Again the corresponding spin two deformations with dimension $6$ can be set to an arbitrary value by an appropriate choice of the homogeneous solution for $\Omega$.

\section{Discussion}
\label{Conclusions}

We have obtained a number of solutions to the type IIB supergravity equations of motion which are deformations of AdS$_3 \times S^3 \times T^4$ and whose non-compact part retains the symmetries of a Schr\"odinger algebra. The dynamical exponent $n$ featuring in our solutions can take integer or half-integer values. Our solutions are classified by an arbitrary vector harmonic $\mathcal A$ on the $S^3$ with eigenvalue $2-n^2$ (which can be nonzero only for integer $n$) plus an arbitrary $S^3$ scalar harmonic $\Omega_{\text h}$ with eigenvalue $-4n(n-1)$. In the dual field theory these deformations correspond to switching on a source for irrelevant vector and spin two operators, respectively. For $n=2$ and/or $\mathcal A = 0$ a subset of our solutions retains four supersymmetries whereas in the other cases we expect that no supersymmetry is preserved.   

There are several interesting avenues which deserve further study. First of all it would be interesting to understand whether our solutions are stable and geodesically complete. In general the function $\Omega$ will take positive and negative values on $S^3$ since it is a linear combination of scalar spherical harmonics and as discussed in \cite{Hartnoll:2008rs} this can be problematic. In fact, given their ubiquitous presence as solutions to various supergravity theories, it will be interesting to investigate further the general question of stability for Schr\"odinger space-times, especially the ones that preserve some supersymmetry. 

There are of course also other supergravity solutions with an $AdS_3$ factor in the metric. It will be interesting to understand whether they also admit Schr\"odinger deformation and to study the spectrum of allowed dynamical exponents. For example, one can study deformations similar to the ones in \eqref{Ansatz1}-\eqref{Ansatz3} for the near horizon limit of a black string in five-dimensional ungauged supergravity or for supersymmetric compactifications of M-theory of the form $AdS_3\times S^2\times CY_3$.

From the perspective of the dual field theory, we have demonstrated that the deformed spacetimes correspond to switching on sources for irrelevant operators. Generally, such an irrelevant deformation induces an infinite number of other deformations via counterterms and in this light the simplicity of the dual gravity solution is rather remarkable. There are several avenues one may pursue to better understand the structure of the dual field theory. For example, one may try to generalize the non-renormalization arguments presented in \cite{Guica:2010sw} to include the mixing with spin two operators. Secondly, it would be interesting to perform an analysis along the lines of \cite{balt} to these spacetimes and investigate whether the field theory is also deformed by multi-trace operators. (Such deformations can in principle have arbitrarily high spin and there may be infinitely many of them, even for sources generated by a null vector.) Lastly, some of the $n=2$ solutions can be obtained by a null Melvin twist and these might correspond to a `dipole' version of the D1-D5 CFT, in analogy with the $n=2$ deformations of $AdS_5\times S^5$ which are dual to `dipole deformations' of $\mathcal{N}=4$ SYM. 

The supersymmetric solutions we have found can be further analyzed in various ways. First of all, it would be interesting to understand if we can find additional supersymmetries for some of our backgrounds along the lines of \cite{Donos:2009zf}. Second, one may compute the superisometry algebra of these solutions and study the corresponding super-Schr\"odinger algebra in 1+1 dimensions along the lines of \cite{Sakaguchi:2008rx}. It will also be interesting to understand whether our supersymmetric solutions admit some super-coset description \cite{SchaferNameki:2009xr}. Furthermore, in \cite{Gutowski:2003rg} supersymmetric solutions of six-dimensional ungauged supergravity were classified and it would be interesting to understand how our solutions fit into this classification. From the point of view of the dual field theory, the supersymmetric solutions suggest that there are supersymmetric deformations of the dual 2D CFT which break Lorentz invariance. It will be interesting to understand these operators better and see if there is some ``chiral ring" structure. 

In the cases where $\mathcal A = 0$ but $\Omega$ is non-vanishing we can arrange for the background \eqref{Ansatz1}-\eqref{Ansatz3} to be purely of NS form. From the viewpoint of the worldsheet CFT these backgrounds correspond to deformations of the well-known WZW model that describes $AdS_3\times S^3\times T^4$ with purely NS flux. It would be interesting to see whether some of these solutions correspond to new exact backgrounds of string theory. This would imply that the corresponding WZW deformations are exactly marginal from the worldsheet point of view and will be similar to the exactly marginal deformations reviewed in \cite{Orlando:2006cc}.

\bigskip
\bigskip
\bigskip
\leftline{\bf Acknowledgements}
\smallskip
  
We would like to thank Mike Douglas and Kostas Skenderis for useful discussions. NB would like to thank Arnab Kundu and Krzysztof Pilch for the enjoyable collaboration on \cite{Bobev:2009mw} and for numerous illuminating conversations. 




\appendix

\section{Killing spinors on $AdS_3\times S^3 \times T^4$}
\renewcommand{\theequation}{A.\arabic{equation}}
\setcounter{equation}{0} 
\label{appendixA}

In this appendix we find the explicit Killing spinors which solve the supersymmetry variations for the $AdS_3\times S^3\times T^4$ background of IIB supergravity. The Killing spinor is given by $\epsilon=e^{i\chi}(\epsilon_1 + i\epsilon_2)$ where
\begin{eqnarray}
\epsilon_1 &=& e^{\frac{\theta}{2}\Gamma^{56}} e^{\frac{(\phi_1 + \phi_2)}{2}\Gamma^{45}} r^{\frac{1}{2}\Gamma^{12}} e^{-v (\Gamma^1-\Gamma^2)\Gamma^3}  \zeta_1^{(0)}~, \label{eps1AdS3S3}\\
\epsilon_2 &=& e^{-\frac{\theta}{2}\Gamma^{56}} e^{\frac{(\phi_1 - \phi_2)}{2}\Gamma^{45}}  r^{-\frac{1}{2}\Gamma^{12}} e^{-\frac{u}{2}(\Gamma^1+\Gamma^2)\Gamma^3} \zeta_2^{(0)}~,\label{eps2AdS3S3}
\end{eqnarray}
and $\zeta_1^{(0)}$ and $\zeta_2^{(0)}$ are constant, real spinors with 32 components which satisfy the projection conditions  
\begin{equation}
(1+\Gamma^{11})\zeta_1^{(0)}=(1+\Gamma^{11})\zeta_2^{(0)}=0~, \qquad\qquad  (1+\Gamma^{123456})\zeta_1^{(0)}=(1+\Gamma^{123456})\zeta_2^{(0)}=0~. \label{projectorsAdSS3}
\end{equation}
To derive \eqref{eps1AdS3S3} and \eqref{eps2AdS3S3} we have also used the following identities
\begin{equation}
e^{\xi \,\Gamma^{MN}} = \cos\xi + \sin\xi ~\Gamma^{MN}~, \qquad~~ e^{\xi \,\Gamma^{1N}} = \cosh\xi + \sinh\xi~\Gamma^{1N}~, \qquad M,N>1~,~~~ \xi \in \mathbb{R}~.
\end{equation}
Each of the two projectors in \eqref{projectorsAdSS3} reduces the components of $\zeta_1^{0}$ and $\zeta_2^{0}$ by half so at the end we have eight independent components in $\zeta_1^{0}$ and eight more in $\zeta_2^{0}$. Therefore we have reproduced the well known result that the $AdS_3\times S^3\times T^4$ solutions of IIB supergravity is 1/2 BPS, \emph{i.e.} it preserves 16 supersymmetries.



\end{document}